# A First-Principles Investigation of Goldene for Enhanced Hydrogen Evolution Reaction

Ashutosh Krishna Amaram, Aaditya Roy and Raghavan Ranganathan*

Department of Materials Engineering, Indian Institute of Technology Gandhinagar, Palaj, Gujarat, 382355, India

* Corresponding author: rraghav@iitgn.ac.in

## Abstract

The recent synthesis of Goldene, a 2D sheet of gold exfoliated from $Ti_3AuC_2$, offers high specific surface area (~260 $m^2g^{-1}$), roughly twice that of fine nanodots (~100 $m^2g^{-1}$), and unique electronic properties due to its dense d-orbital. In this work, we investigate the adsorption of single atom catalyst (SAC) of hydrogen atom on pristine goldene (pG), monovacant goldene (vG), and sulfur-functionalized variants (thiol-pG and thiol-vG) using ab initio calculations. The adsorption energy of a single H atom and $\Delta G_H$, determines the efficiency of the Volmer step of the hydrogen evolution reaction (HER) and is a key descriptor for HER activity. We explore various potential sites for H adsorption and its impact on descriptors such as Bader charges, d-band shift and the exchange current density.

**Keywords:** Density Functional Theory (DFT); Hydrogen Evolution Reaction (HER); 2D Materials; Goldene

## 1. INTRODUCTION

The successful isolation of graphene in 2004[1] has ushered in a new era of research on two-dimensional (2D) materials, with significant advancements and commercial deployment over the past decade. Hexagonal Boron Nitride (h-BNs), transition metal dichalcogenides (TMDs) such as $MoS_2$, MXenes and several other 2D materials have found immense potential for a variety of applications such as in transistors,[2] memristors,[3] batteries,[4,5] sensing devices,[6,7] catalysis,[8] selective adsorption,[9] etc. This class of materials shows unique properties which otherwise is not seen in bulk materials. For instance, graphene shows exceptionally high electron mobility of 200,000 cm$^2$/Vs,[10] with a theoretically high specific surface area of ~2,600 m$^2$/g,[11] and a very small pore size that allows it to be used or a variety of electrocatalysis sensing and filtration applications.[12] Graphene doped with nitrogen also shifts the hydrogen adsorption, measured by the adsorption free energy change, $\Delta G_H$ closer to thermoneutral values.[13] TMDs such as $MoS_2$ and $WS_2$ shows metallic edge catalysis, as they are more active than the basal planes,[14] particularly for $CO_2$ reduction, and, in the presence of sulphur vacancy, the activation barrier is lowered and forms stabilized *COOH intermediates.[15] In general, 2D materials possess high specific surface area along with vacancies, exposed active sites (edges, vacancies), and tunable electronic structures that enhance their catalytic performance. Recently, goldene[2], a freestanding 2D sheet of gold, exfoliated from $Ti_3AuC_2$ by selectively etching out the MXene layers was synthesized. It has a hexagonal lattice in the P6/mmm space group.[16,17,18] Comprehensive studies of goldene using density functional theory (DFT) and machine learning interatomic potentials show that it is highly stable, exhibiting strong mechanical characteristics such as elastic moduli of approximately 226 GPa and tensile strength of up to 12 GPa.[19] Due to its metallic nature, it is expected to have exceptional electron mobility. It was also theoretically shown to be thermally stable up to 1400 K with a notably low lattice thermal conductivity of around 10-12 W/m.K and an intrinsic conductivity of approximately 2.5x10$^6$ S/m at room temperature, which is attributed to its dense d-orbital population and its lattice symmetry.[20,21] The conductivity is comparable to that of bulk gold, making it a potential candidate for high performance electronic devices. Despite these promising characteristics, the catalytic properties of goldene, particularly in the context of the hydrogen evolution reaction, remains underexplored. Previous studies on 2D Au (111) surfaces have reported moderate HER but did not examine the effects of atomic vacancies or surface functionalization.[22] The present study explores the HER on goldene surface, particularly the Volmer step on 4 different surfaces of goldene: pristine goldene (pG), mono-vacant goldene (vG), thiol functionalized goldene and mono-vacant thiol functionalized goldene (thiol-pG/vG). HER is pivotal for green hydrogen production enabling sustainable energy storage through water electrolysis. In acidic media, HER follows a two-electron transfer mechanism involving three steps – Volmer step[23,24], Heyrovsky step, and Tafel step. The Volmer step involves the electrochemical adsorption of hydrogen (H⁺ + e⁻ → H*) involving a proton coupled electron transfer (PCET), where the hydrogen adsorption energy, $\Delta G_H$ determines the reaction rate. In the H$_2$O + e⁻ + M → M-H + OH⁻ Volmer step a catalyst is required to facilitate the cleavage of the H-O-/-H bond in water This additional water dissociation step in alkaline medium makes the process more sluggish compared to acidic media. The kinetics of the Volmer step are sensitive to the catalyst's ability to adsorb and activate the protons or water where the local electrostatic potential, surface structure and composition of the catalyst plays a very important role.[25] For example, catalysts with optimal hydrogen binding energy ($\Delta G_H \sim 0$) along with efficient water dissociation capabilities show significant acceleration in the Volmer step and thus enhance the overall HER activity. The adsorption energy $E_{ads}$ is given by equation (1).

$$E_{ads} = E_{\{surface+H\}} - E_{surface} - \frac{1}{2}E_{H^2} \quad (1)$$

where, $E_{\{surface+H\}}$ is the energy of surface when H has been adsorbed. The relation between $\Delta G_H$ and $E_{ads}$ is given by equation (2).

$$\Delta G = E_{ads} + \Delta E_{ZPE} - T\Delta S \quad (2)$$

Here, $\Delta E_{ZPE}$ is the difference in zero-point energies between the adsorbed state and the gas phase, and $T\Delta S$ is the entropy correction. At 298 K, substituting the values and we get the equation[26] as,

$$\Delta G = E_{ads} + 0.24 \; eV \quad (3)$$

To analyze the electron transfer mechanisms, two critical electronic descriptors, namely Bader 4charge analysis and d-band center shifts were carried out. Bader charge analysis quantifies the charge transfer between adsorbates and substrates, revealing how electron redistribution modulates the adsorption strength. For instance, a greater transfer of electron density to the adsorbed hydrogen (H*) typically weakens the H–surface bond, potentially bringing the hydrogen adsorption free energy ($\Delta G_H$) closer to the ideal value near zero, which is optimal for HER. Complementing this, the d-band center provides a mechanistic descriptor for the strength of adsorbate–surface interactions. The position of the d-band center relative to the Fermi level serves as a predictor for hydrogen binding: a d-band center closer to the Fermi level generally results in stronger adsorption, while a d-band center away from the Fermi level value weakens hydrogen binding. Surface modifications such as introducing mono-vacancies or attaching thiol groups can shift the d-band center, thereby fine-tuning the adsorption energy and catalytic activity.[28]

## 2. METHODOLOGY

Density functional theory (DFT) calculations were performed using the Vienna Ab Initio Simulation Package (VASP) with the Projector Augmented Wave (PAW) method with the Perdew-Burke-Ernzerhof (PBE) exchange-correlation functional.[29,30] Two primary systems were modeled: pristine goldene (a 2D Au monolayer) with 16 atoms, and sulphur-doped goldene with and without vacancies. The 16-atom system was selected based on the convergence of the energy in the vacancy system from 4, 9, 16, to 25 atoms. Structural relaxation was performed using the conjugate-gradient algorithm with a plane wave cutoff energy of 530 eV. Gaussian smearing (with $\sigma = 0.01$ eV), and an electronic convergence of $1\times10^{-6}$ eV were used. The atomic positions were optimized over a maximum 100 ionic steps while maintaining fixed lattice parameters. The H-atom adsorption was carried out at 3 sites, on top of Au/vacancy, bridge site between Au atoms, and above the triangular pore. Calculations without spin-polarization were performed, with convergence of $10^{-4}$ eV for geometry optimization during adsorption.

## 3. RESULTS AND DISCUSSIONS

We first begin our study by identifying different possible sites of adsorption of hydrogen atom on pG as shown in Figure 1. The first site of adsorption is right above a gold atom. At this site we observed a very high adsorption energy ($E_{ads}$) of 1.037 eV which led to a very high positive value of $\Delta G_H$ (as seen in Figure 5(a)), overall making this site unsuitable for effective HER. Moving on to the next site, the hydrogen atom is made to adsorb between two gold atoms (right above the gold-gold bond) (as shown in Figure 1(b, e)). At this site we see a considerable decrease to 0.364 eV in $E_{ads}$ with a $\Delta G_H$ of 0.604 eV (Figure 5(a)) which is still not the best compared to traditional transition metals like Pt and Pd. The last and final site (as shown in Figure 1(c, f)) explored for pG was just above the triangular void observed in Goldene sheets. This was the best site for pG by achieving an $E_{ads}$ value of 0.224 eV and a $\Delta G_H$ of 0.464 eV (as seen in Figure 5(a)), making this site a potential candidate for HER. However, this site still does not stand close to the values achieved by Pd and Pt, hence the creation and study of vacant sites in these 2D metal sheet becomes important. Vacancies generate unsaturated coordination sites, which serve as highly reactive centers for the adsorption and activation of reactant molecules.

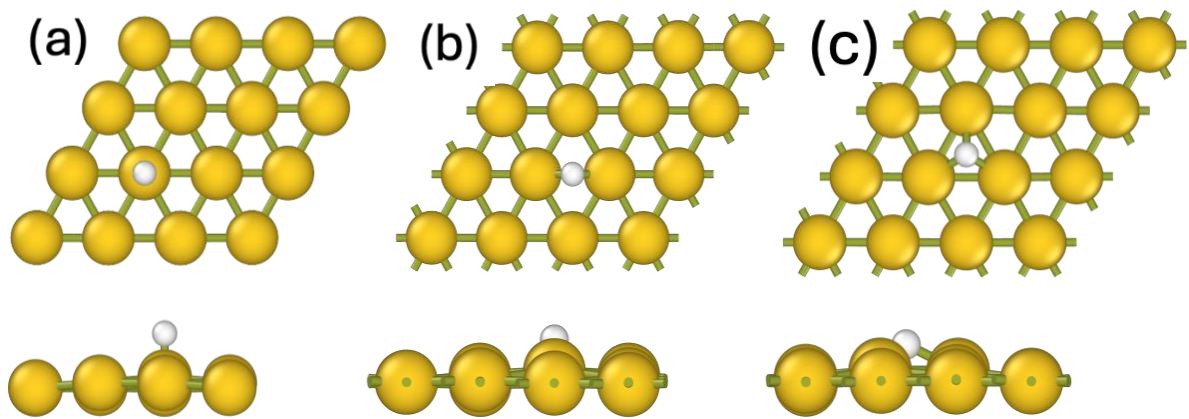

**Figure 1**. (a-c) Top view of site-1, site-2 and site-3 of pristine goldene respectively, (d-f) Side view of site-1, site-2, site-3 of pristine goldene respectively.

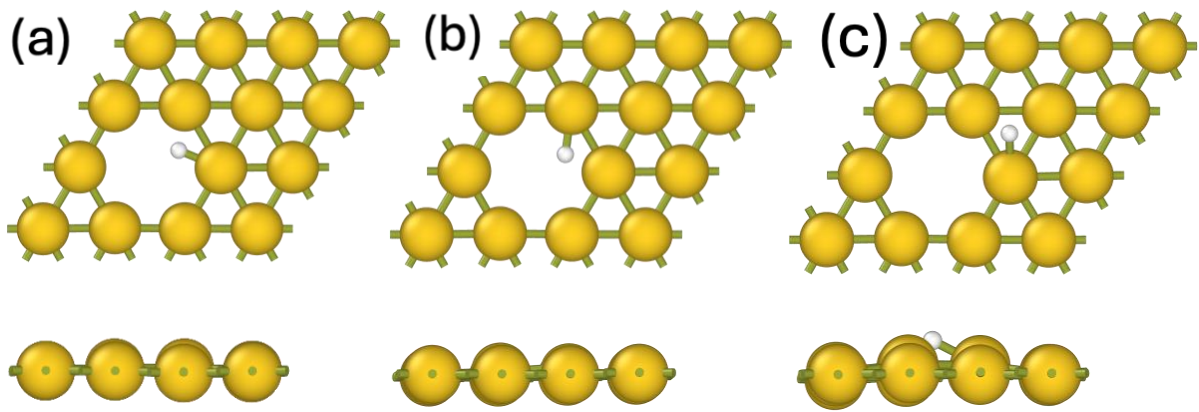

**Figure 2**. (a-c) Top view of site-1, site-2 and site-3 of goldene with a vacancy respectively, (d-f) Side view of site-1, site-2, site-3 of goldene with a vacancy respectively.

At the first two sites, both located adjacent to the vacancy (as seen in Figure 3(a-b)), the hydrogen adsorption energies ($E_{ads}$) were found to be –0.229 eV and –0.228 eV, respectively. These values correspond to nearly thermoneutral Gibbs free energy changes ($\Delta G_H \approx 0.01$ eV, as seen in Figure 5(b)), which is considered optimal for HER, as it ensures both efficient hydrogen adsorption and facile desorption. The third site, positioned further from the vacancy (as seen in Figure 3(c)), exhibited a slightly higher $E_{ads}$ of 0.183 eV, resulting in a $\Delta G_H$ of 0.423 eV (Figure 5(b)). While this is less favourable than the vacancy-adjacent sites, it still represents a significant improvement over the pG surface.

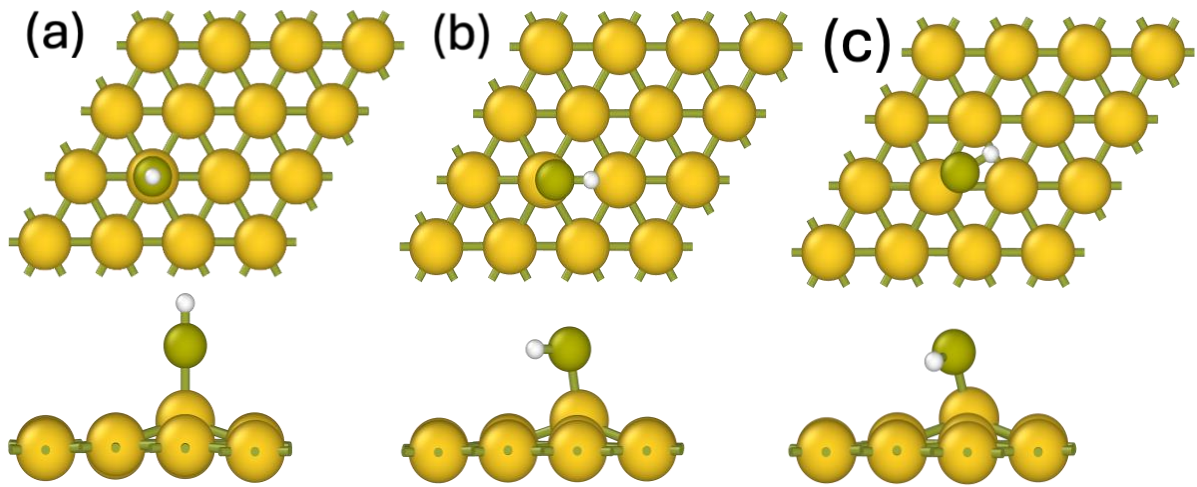

**Figure 3.** (a-c) Top view of site-1, site-2 and site-3 of pristine goldene with thiol functionalization (thiol-pG) respectively, (d-f) Side view of site-1, site-2, site-3 of pristine goldene with thiol functionalization respectively.

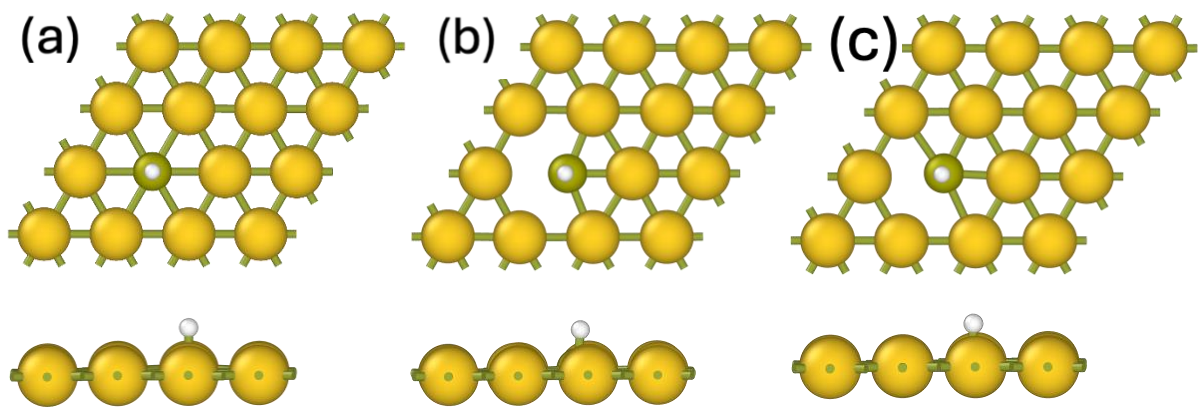

**Figure 4.** (a-c) Top view of site-1, site-2 and site-3 of goldene with a vacancy and thiol functionalization (thiol-vG) respectively, (d-f) Side view of site-1, site-2, site-3 of goldene with a vacancy and thiol functionalization respectively.

On thiol-pG, hydrogen adsorption was markedly enhanced, with $E_{ads}$ values ranging from –0.473 eV to –1.397 eV across the three sites. The corresponding $\Delta G_H$ values were strongly negative, particularly at sites 2 and 3 (–1.107 eV and –1.157 eV, Figure 5(c)), indicating very strong hydrogen binding. While such strong adsorption can increase the density of active sites, it may also hinder hydrogen desorption, potentially limiting the overall HER efficiency. In contrast, thiol-vG surfaces (see Figure 4 for optimized adsorbed snapshots) achieved a more balanced adsorption profile, with $E_{ads}$ values between –0.358 eV and –0.387 eV and $\Delta G_H$ values in the range of –0.118 eV to –0.147 eV (as seen in Figure 5(d)). The data for adsorption characteristics for all the sites explored are presented in Table 1.

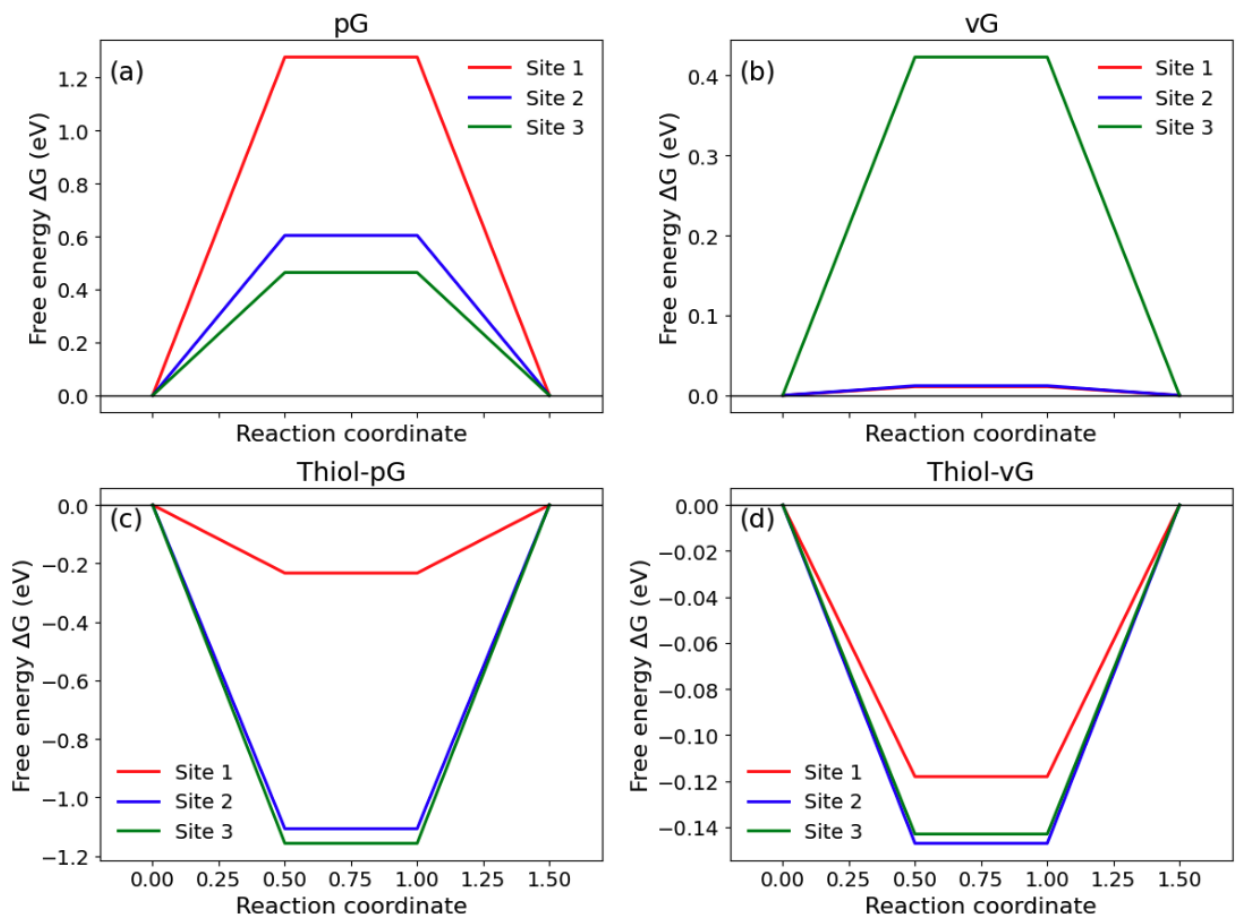

**Figure 5.** Free energy profiles for H adsorption at three sites on (a) pristine goldene (pG), monovacant goldene (vG), (b) thiol functionalized pristine goldene (thiol-pG) and thiol functionalized monovacant goldene (thiol-vG).

**Table 1**. E$_{ads}$ and ΔG$_H$ values at different surfaces of pristine goldene, monovacant goldene, thiol functionalized pristine goldene and thiol functionalized monovacant goldene.

| Surface | Site | E$_{ads}$ (eV) | ΔG$_H$ (eV) |
|---|---|---|---|
| pG | 1 | 1.037 | 1.277 |
|  | 2 | 0.364 | 0.604 |
|  | 3 | 0.224 | 0.464 |
| vG | 1 | −0.229 | 0.011 |
|  | 2 | −0.228 | 0.012 |
|  | 3 | 0.183 | 0.423 |
| Thiol-pG | 1 | −0.473 | −0.233 |
|  | 2 | −1.347 | −1.107 |
|  | 3 | −1.397 | −1.157 |
| Thiol-vG | 1 | −0.358 | −0.118 |
|  | 2 | −0.387 | −0.147 |
|  | 3 | −0.383 | −0.143 |

To map the electron distribution and charge transfer characteristics, Bader charge analysis was performed using Henkelman Group method.[31-34] The net atomic charge, Q$_{Bader}$, for each atom is determined by subtracting the Bader charge from the number of valence electrons, Z$_{val}$, for that atom:

$$Q_{Bader} = Z_{val} - q_{Bader} \qquad (4)$$

Where, Q$_{Bader}$ is the number of electrons assigned to the atom by Bader analysis. As seen in Figure 6(a), for pG, Bader charges on Au atoms at all three hydrogen adsorption sites (site 1: atop, site 2: bridge, site 3: hollow) remain close to neutral, typically within ±0.03 electrons. This indicates minimal charge redistribution upon hydrogen adsorption, thus retaining the metallic character of the goldene. As seen in Figure 6(b), the creation of a vacancy induces slightly higher charge deviations (up to ±0.05 electrons) for gold atoms near the vacancy, reflecting localised electronic perturbation, but the overall charge transfer remains meagre. From Figure 6(c,d) its observed that thiol functionalization of goldene sheets leads to a significant change in Bader charges especially at the active sites: the sulphur atom exhibits a large positive Bader charge (+1.02 to +1.30 electrons), indicating substantial electron loss whereas the hydrogen atom shows a significant negative Bader charge (–1.18 to –1.32 electrons), reflecting strong electron gain. The gold atoms near the functional group display slightly higher charge deviations (up to ±0.09 electrons), particularly for atoms directly bonded to sulphur or adjacent to the vacancy. This shows that the introduction of the thiol group to goldene dramatically enhances charge transfer between sulphur and hydrogen. Previous research has shown that electron-enriched sites typically show enhanced HER activity as electron-rich sites are often more reactive, facilitating the activation and reduction of protons, thereby promoting $\Delta G_H$ to thermoneutral values.[35,36]

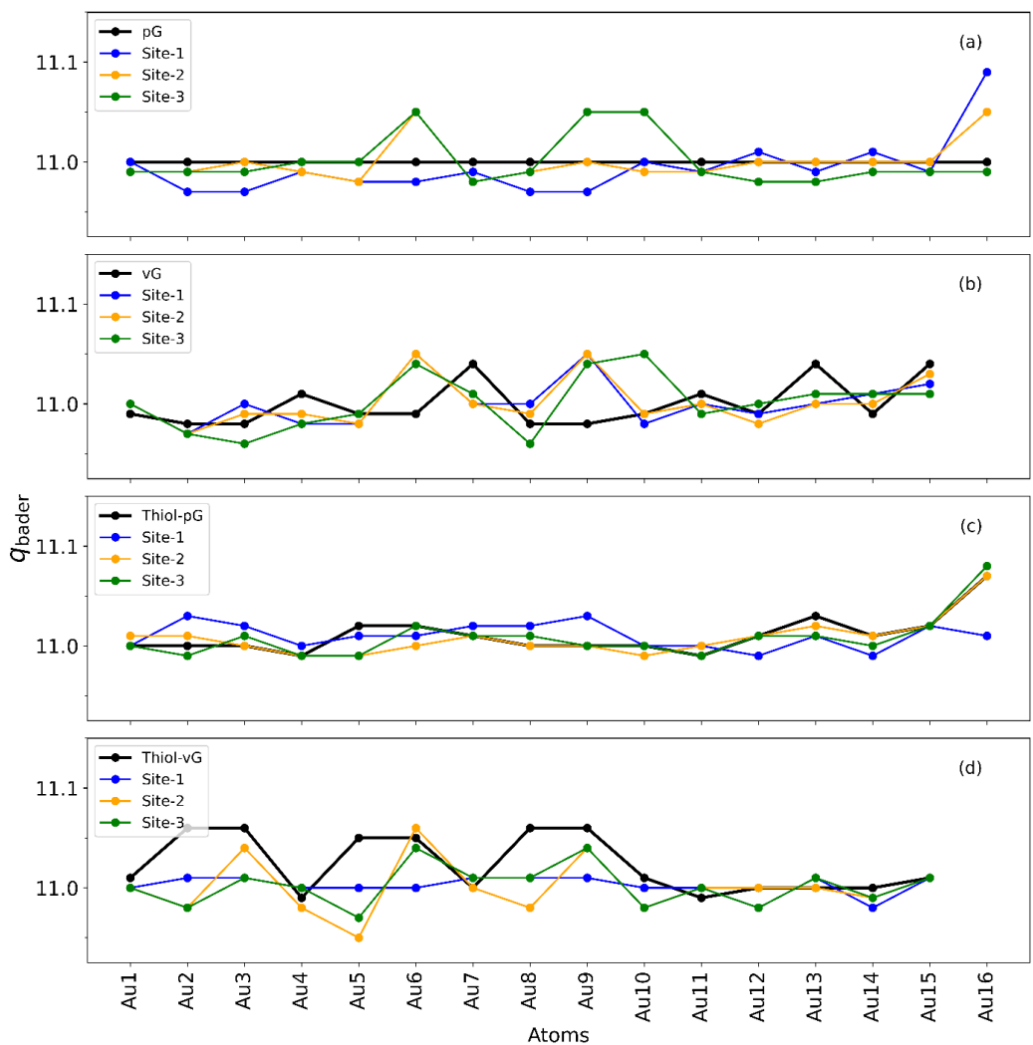

**Figure 6.** Bader Charge of gold atoms in **(a)** pristine goldene, **(b)** monovacant goldene, **(c)** thiol functionalized pristine goldene **(d)** thiol functionalized monovacant goldene.

Another important aspect studied was the d-band center first proposed by Hammer et al.[37] In pG, the d band center, $E_d$ values are -2.76 eV, -2.74 eV, and -2.73 eV for sites 1, 3, and 2, respectively, indicating relatively deep d-states and moderate hydrogen adsorption strength. Upon introducing a monovacancy, these values shift upward to -2.62 eV (site 1), -2.60 eV (site 3), and -2.62 eV (site 2), representing a notable upshift of ~0.12 eV across all active sites. This upward shift in $E_d$ correlates with stronger hydrogen adsorption, as predicted by d-band theory, and aligns with Bader charge analysis of enhanced charge transfer near vacancy sites. The hollow site (site 3) in vG retains the highest $E_d$ (-2.60 eV), which is indicative of its favorable orbital interaction with hydrogen and superior HER performance. In comparison, thiol-functionalized systems exhibit a consistent downshift in d-band centers relative to both pG and vG. For the thiol-pG system, $E_d$ values are -2.60 eV (site 1), -2.62 eV (site 3), and -2.71 eV (site 2), while the thiol-vG system shows even deeper d-band centers at -2.74 eV, -2.77 eV, and -2.78 eV at the respective sites. These shifts represent a downshift of approximately 0.14 to 0.18 eV compared to thiol-pG, suggesting a weakening of hydrogen adsorption strength. This is attributed to electron donation from the sulfur atoms in the thiol groups to the gold surface, which lowers the d-band center and reduces occupancy of antibonding states. Mechanistically, this behavior is explained by the relation $E_{ads} \propto V_{ad}^2/\epsilon_d-\epsilon_a$, where a lower $E_d$ reduces hydrogen binding energy and thus promotes hydrogen desorption—a favorable trait for HER kinetics. Among the thiolated systems, the hollow site (site 3) again stands out with the highest $E_d$ (-2.62 eV in thiol-pG), confirming its consistent role as the most HER-active site across all systems.

The catalytic performance of goldene can be further explored with the help of partial density of states plots (PDOS) as shown in Figure 7 and Figure 8. For site-1 of pG, we had observed a high $\Delta G_H$ value of 1.227 eV. This value can be justified by observing Figure 7(a) and Figure 8(a) where for this particular site, we observe additional conduction states spread across the energy spectrum. Additionally, this observation is further accompanied by a sudden increase in the number of available conduction states in Hydrogen atom. For vG, which yielded the best results for HER, we do not observe much change in DOS of both goldene and hydrogen atom as seen in Figure 7(b) and Figure 8(b). For thiol functionalized pristine and monovacant goldene sheets, the DOS of goldene more or less stays the same as seen in Figure 7(c,d) but in case of Hydrogen atoms we see some additional states arising (as seen in Figure 8(c,d)) which is likely due to the introduction of a thiol functional group.

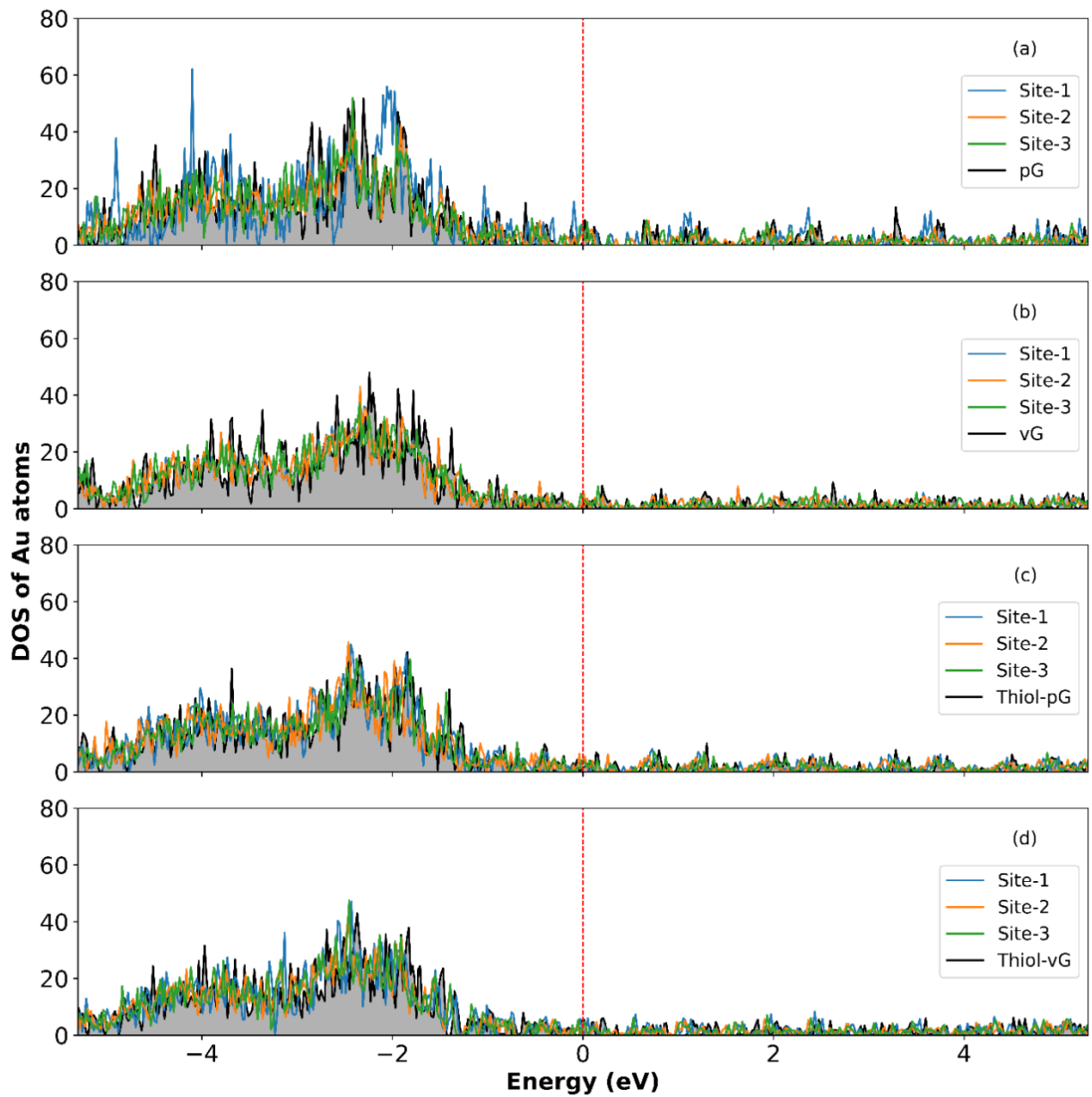

**Figure 7.** Density of states of gold atoms (eV) in **(a)** pristine goldene, **(b)** monovacant goldene, **(c)** thiol functionalized pristine goldene, **(d)** thiol functionalized monovacant goldene.

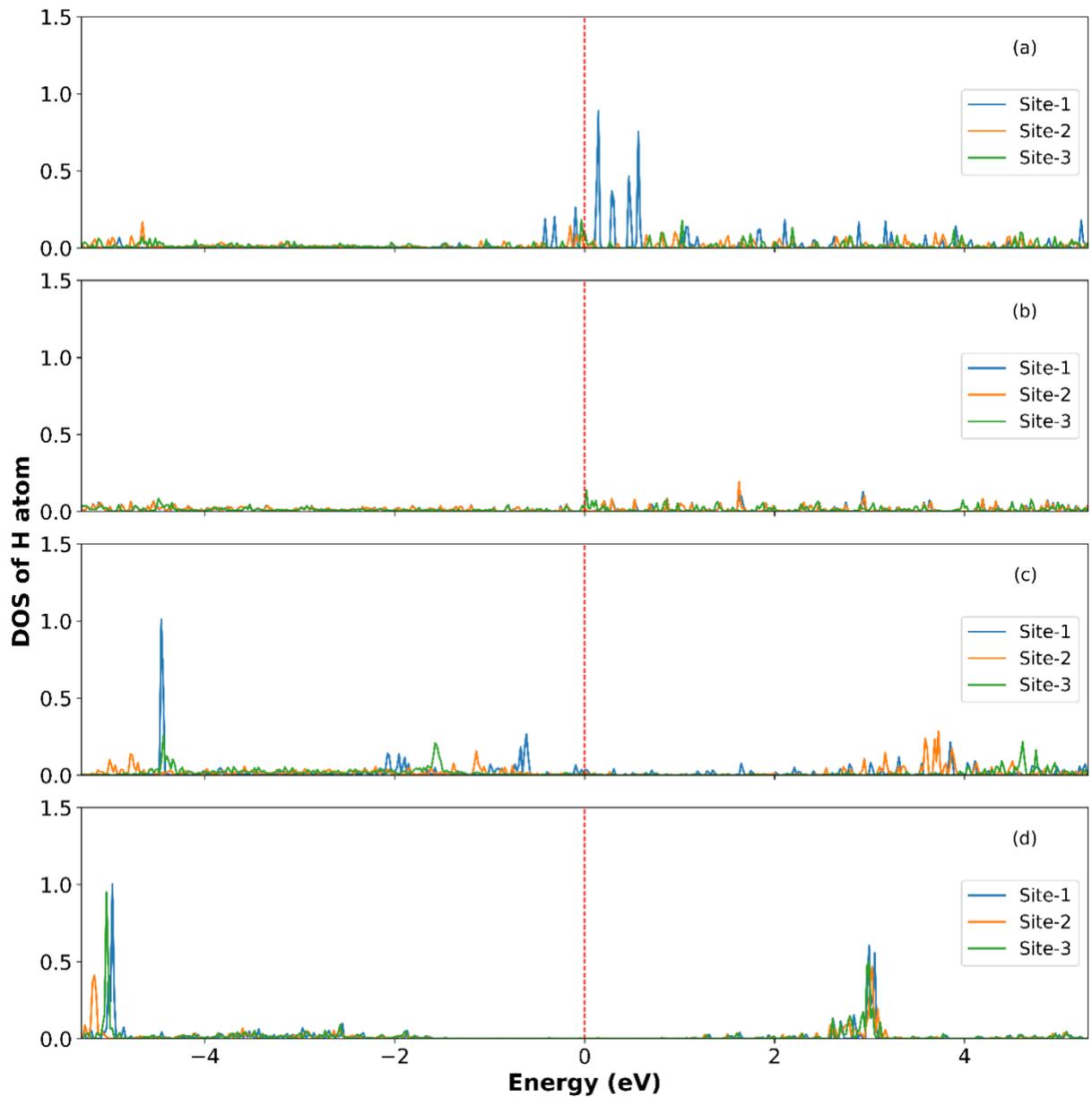

**Figure 8.** Density of states of hydrogen atom (eV) in **(a)** pristine goldene, **(b)** monovacant goldene, **(c)** thiol functionalized pristine goldene, **(d)** thiol functionalized monovacant goldene.

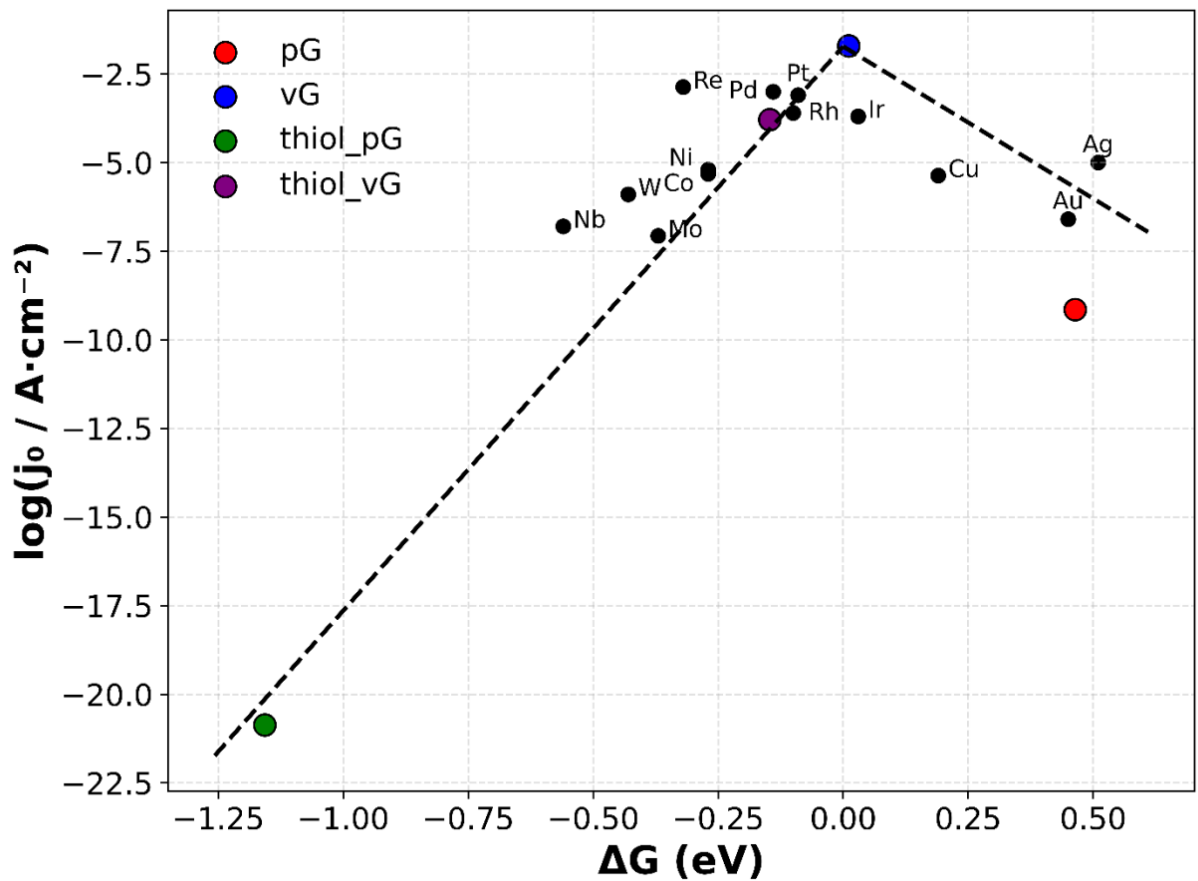

**Figure 9.** Volcano plot depicting the relationship between hydrogen adsorption energy ($\Delta G_H$) and exchange current density ($\log(j_o)$) for reference metals[27] and 2D goldene systems.

The volcano plot (Figure 9) illustrates the well-established Sabatier-type relationship between hydrogen adsorption free energy ($\Delta G_H$) and catalytic activity ($\log_{10} j_o$) for the hydrogen evolution reaction. The two dashed black regression lines represent the strong-binding (left) and weak-binding (right) branches, intersecting at the volcano apex ($\Delta G_H \approx 0$ eV). Catalysts positioned near this apex balance adsorption and desorption steps, hence achieving optimal HER kinetics. In our case, we have adopted the original kinetic prefactor $k_o = 200$ s$^{-1}$·site$^{-1}$, consistent with foundational studies by Nørskov et al,[27] but may vary in experimental analysis.

The reference metal data align on the expected volcano trend, validating our approach. Our 2D goldene-derived systems—pristine pG and vG, and thiol-functionalized analogues—are plotted alongside. Notably, the thiol-vG system appears closest to the volcano apex, suggesting the most favourable HER activity among the series. However, although thiol-vG appears to be the best candidate on the plot, such placement must be interpreted cautiously: prior work shows that DFT-based predictions of ΔG$_H$ values carry typical uncertainties of ~±0.2 eV, which can translate into significant shifts along the volcano scale and hence into different predicted activities. For instance, Di Liberto et al. report ±0.2 eV uncertainty in ΔG$_H$ descriptors for single-atom catalysts.[38] Thus, while the trend of weakened hydrogen adsorption for the thiol-functionalized systems (consistent with the down-shifted d-band centre's) is robust, the exact ranking of activity should be treated with due caution.

The application of conventional volcano-plot models derived from bulk-metal kinetics[27] to 2D materials presents fundamental limitations. The fitted kinetic prefactor obtained in our analysis yielded an effective $k_0 = 3.53 \times 10^3 \text{ s}^{-1} \text{ site}^{-1}$ which deviates substantially from the standard value $k_0 = 200 \text{ s}^{-1} \text{ site}^{-1}$ used for bulk surfaces. This discrepancy indicates that the traditional Nørskov-type kinetic treatment may not accurately capture reaction dynamics on low-dimensional catalysts, where quantum confinement, reduced coordination, and altered electronic screening strongly affect charge transfer kinetics.

## 4. CONCLUSION

In this work, we carried out an investigation on the catalytic behaviour of goldene and demonstrate that it is highly sensitive to atomic-scale modifications. While pristine goldene exhibits moderate hydrogen adsorption energetics, introducing gold vacancies significantly enhances its catalytic performance, yielding near-thermoneutral Gibbs free energies of hydrogen adsorption, ΔG$_H$, at vacancy-adjacent sites. Further improvement is achieved through thiol functionalization, which alters the local charge environment and shifts the d-band centre—fine-tuning hydrogen binding strength to promote efficient HER activity in select configurations.

Analysis of the local electronic structure, including Bader charge distributions and d-band centre shifts, reveals that vacancy sites enhance local electron density, facilitating proton activation, whereas sulphur functionalization can generate both optimal (ΔG$_H$ ≈ 0 eV) and overly strong hydrogen-binding sites. These electronic descriptors provide a clear, atomistic rationale for the observed variations in HER activity across different goldene structures. A corresponding volcano plot analysis places the most active configurations of goldene on par with established transition-metal catalysts in terms of the hydrogen adsorption descriptor.

These findings highlight that HER catalysis on unconventional 2D surfaces, especially those with low coordination or single-atom features should be regarded as an emerging field where existing bulk-derived descriptors may require significant re-evaluation. However these 2D metal sheets provide exciting horizons for catalysis especially for HER which is key to move forward in realising the possibility of having a clean fuel that utilises HER for constant hydrogen supply in Hydrogen Fuel Cells.

## ACKNOWLEDGEMENT

The authors acknowledge the use of Param Ananta supercomputing facility of the Indian Institute of Technology Gandhinagar to carry out all the simulations reported in this work.

## REFERENCES

1. Novoselov, K. S.; Geim, A. K.; Morozov, S. V.; Jiang, D.; Zhang, Y.; Dubonos, S. V.; Grigorieva, I. V.; Firsov, A. A. Electric Field Effect in Atomically Thin Carbon Films. *Science* **2004**, *306* (5696), 666–669. https://doi.org/10.1126/science.1102896.

2. Laturia, A.; Van de Put, M. L.; Vandenberghe, W. G. Dielectric Properties of Hexagonal Boron Nitride and Transition Metal Dichalcogenides: From Monolayer to Bulk. *npj 2D Mater. Appl.* **2018**, *2*, 6. https://doi.org/10.1038/s41699-018-0050-x.

3. Zhang, S.; Tao, Y.; Qin, S. et al. Memristors Based on Two-Dimensional h-BN Materials: Synthesis, Mechanism, Optimization and Application. *npj 2D Mater. Appl.* **2024**, *8*, 81. https://doi.org/10.1038/s41699-024-00519-z.

4. Ismail, K. B. M.; Kumar, A.; Mahalingam, S.; Kim, J.; Atchudan, R. Recent Advances in Molybdenum Disulfide and Its Nanocomposites for Energy Applications: Challenges and Development. *Materials (Basel)* **2023**, *16* (12), 4471. https://doi.org/10.3390/ma16124471.

5. Askari, M. B.; Salarizadeh, P.; Veisi, P.; Samiei, E.; Saeidfirozeh, H.; Tourchi Moghadam, M. T.; Di Bartolomeo, A. Transition-Metal Dichalcogenides in Electrochemical Batteries and Solar Cells. *Micromachines* **2023**, *14*, 691. https://doi.org/10.3390/mi14030691.

6. Castelletto, S.; Inam, F. A.; Sato, S. I.; Boretti, A. Hexagonal Boron Nitride: A Review of the Emerging Material Platform for Single-Photon Sources and the Spin-Photon Interface. *Beilstein J. Nanotechnol.* **2020**, *11*, 740–769. https://doi.org/10.3762/bjnano.11.61.

7. Weng, Q.; Wang, X.; Wang, X.; Bando, Y.; Golberg, D. Functionalized Hexagonal Boron Nitride Nanomaterials: Emerging Properties and Applications. *Chem. Soc. Rev.* **2016**, *45*, 3989–4012. https://doi.org/10.1039/C5CS00869G.

8. Saeloo, B.; Saisopa, T.; Chavalekvirat, P.; Iamprasertkun, P.; Jitapunkul, K.; Sirisaksoontorn, W.; Lee, T. R.; Hirunpinyopas, W. Role of Transition Metal Dichalcogenides as a Catalyst Support for Decorating Gold Nanoparticles for Enhanced Hydrogen Evolution Reaction. *Inorg. Chem.* **2024**, *63* (40), 18750–18762. https://doi.org/10.1021/acs.inorgchem.4c02668.

9. Lu, Y.; Li, B.; Xu, N. et al. One-Atom-Thick Hexagonal Boron Nitride Co-Catalyst for Enhanced Oxygen Evolution Reactions. *Nat. Commun.* **2023**, *14*, 6965. https://doi.org/10.1038/s41467-023-42696-3.

10. Lee, J.-H.; Park, S.-J.; Choi, J.-W. Electrical Property of Graphene and Its Application to Electrochemical Biosensing. *Nanomaterials* **2019**, *9*, 297. https://doi.org/10.3390/nano9020297.

11. Zhang, S.; Wang, H.; Liu, J.; Bao, C. Measuring the Specific Surface Area of Monolayer Graphene Oxide in Water. *Mater. Lett.* **2020**, *261*, 127098. https://doi.org/10.1016/j.matlet.2019.127098.

12. Nair, R. R.; Wu, H. A.; Jayaram, P. N.; Grigorieva, I. V.; Geim, A. K. Unimpeded Permeation of Water Through Helium-Leak–Tight Graphene-Based Membranes. *Science* **2012**, *335*, 442–444. https://doi.org/10.1126/science.1211694.

13. Duan, X.; Ao, Z.; Sun, H.; Indrawirawan, S.; Wang, Y.; Kang, J.; Liang, F.; Zhu, Z. H.; Wang, S. Nitrogen-Doped Graphene for Generation and Evolution of Reactive Radicals by Metal-Free Catalysis. *ACS Appl. Mater. Interfaces* **2015**, *7* (7), 4169–4178. https://doi.org/10.1021/am508416n.

14. Jaramillo, T. F.; Jørgensen, K. P.; Bonde, J.; Nielsen, J. H.; Horch, S.; Chorkendorff, I. Identification of Active Edge Sites for Electrochemical H2 Evolution from MoS2 Nanocatalysts. *Science* **2007**, *317*, 100–102. https://doi.org/10.1126/science.1141483.

15. Kumar, D. R.; Ranjith, K. S.; Manoharan, M.; Haldorai, Y.; Han, Y.-K.; Oh, T. H.; Kumar, R. T. R. Sulfur Vacancies Promoted Highly Efficient Visible Light Photocatalytic Degradation of Antibiotic and Phenolic Pollutants over WS$_2$/rGO Heterostructure. *Sep. Purif. Technol.* **2024**, *329*, 125172. https://doi.org/10.1016/j.seppur.2023.125172.

16. Kashiwaya, S.; Shi, Y.; Lu, J. et al. Synthesis of Goldene Comprising Single-Atom Layer Gold. *Nat. Synth.* **2024**, *3*, 744–751. https://doi.org/10.1038/s44160-024-00518-4.

17. Sharma, S. K.; Pasricha, R.; Weston, J.; Blanton, T.; Jagannathan, R. Synthesis of Self-Assembled Single Atomic Layer Gold Crystals-Goldene. *ACS Appl. Mater. Interfaces* **2022**, *14* (49), 54992–55003. https://doi.org/10.1021/acsami.2c19743.

18. Zhao, S.; Zhang, H.; Zhu, M.; Jiang, L.; Zheng, Y. Electrical Conductivity of Goldene. *Phys. Rev. B* **2024**, *110* (8), 085111. https://doi.org/10.1103/PhysRevB.110.085111.

19. Mortazavi, B. Goldene: An Anisotropic Metallic Monolayer with Remarkable Stability and Rigidity and Low Lattice Thermal Conductivity. *Materials (Basel)* **2024**, *17* (11), 2653. https://doi.org/10.3390/ma17112653.

20. Bhandari, S.; Hao, B.; Waters, K.; Lee, C. H.; Idrobo, J.-C.; Zhang, D.; Pandey, R.; Yap, Y. K. Two-Dimensional Gold Quantum Dots with Tunable Bandgaps. *ACS Nano* **2019**, *13* (4), 4347–4353. https://doi.org/10.1021/acsnano.8b09559.

21. Pifferi, V.; Chan-Thaw, C. E.; Campisi, S.; Testolin, A.; Villa, A.; Falciola, L.; Prati, L. Au-Based Catalysts: Electrochemical Characterization for Structural Insights. *Molecules* **2016**, *21*, 261. https://doi.org/10.3390/molecules21030261.

23. Gutić, S. J.; Dobrota, A. S.; Fako, E.; Skorodumova, N. V.; López, N.; Pašti, I. A. Hydrogen Evolution Reaction-From Single Crystal to Single Atom Catalysts. *Catalysts* **2020**, *10*, 290. https://doi.org/10.3390/catal10030290.

24. Allés, M.; Meng, L.; Beltrán, I.; Fernández, F.; Viñes, F. Atomic Hydrogen Interaction with Transition Metal Surfaces: A High-Throughput Computational Study. *J. Phys. Chem. C* **2024**, *128* (47), 20129–20139. https://doi.org/10.1021/acs.jpcc.4c06194.

25. Santana, J. A.; Meléndez-Rivera, J. Hydrogen Adsorption on Au-Supported Pt and Pd Nanoislands: A Computational Study of Hydrogen Coverage Effects. *J. Phys. Chem. C* **2021**, *125* (9), 5110–5115. https://doi.org/10.1021/acs.jpcc.0c11566.

26. Wilson, J. C.; Caratzoulas, S.; Vlachos, D. G. et al. Insights into Solvent and Surface Charge Effects on Volmer Step Kinetics on Pt (111). *Nat. Commun.* **2023**, *14*, 2384. https://doi.org/10.1038/s41467-023-37935-6.

27. Nørskov, J. K.; Bligaard, T.; Logadottir, A.; Kitchin, J. R.; Chen, J. G.; Pandelov, S.; Stimming, U. Trends in the Exchange Current for Hydrogen Evolution. *J. Electrochem. Soc.* **2005**, *152* (2), J23–J26. https://doi.org/10.1149/1.1856988.

28. Calle-Vallejo, F.; Inoglu, N. G.; Su, H.-Y.; Martínez, J. I.; Man, I. C.; Koper, M. T. M.; Kitchin, J. R.; Rossmeisl, J. Number of Outer Electrons as Descriptor for Adsorption Processes on Transition Metals and Their Oxides. *Chem. Sci.* **2013**, *4* (3), 1245–1249. https://doi.org/10.1039/C2SC21601A.

29. Kresse, G.; Furthmüller, J. Efficient Iterative Schemes for Ab Initio Total-Energy Calculations Using a Plane-Wave Basis Set. *Phys. Rev. B* **1996**, *54* (16), 11169. https://doi.org/10.1103/PhysRevB.54.11169.

30. Kresse, G.; Joubert, D. From Ultrasoft Pseudopotentials to the Projector Augmented-Wave Method. *Phys. Rev. B* **1999**, *59* (3), 1758. https://doi.org/10.1103/PhysRevB.59.1758.

31. Tang, W.; Sanville, E.; Henkelman, G. A Grid-Based Bader Analysis Algorithm without Lattice Bias. *J. Phys.: Condens. Matter* **2009**, *21* (8), 084204. https://doi.org/10.1088/0953-8984/21/8/084204.

32. Sanville, E.; Kenny, S. D.; Smith, R.; Henkelman, G. An Improved Grid-Based Algorithm for Bader Charge Allocation. *J. Comput. Chem.* **2007**, *28* (5), 899–908. https://doi.org/10.1002/jcc.20575.

33. Henkelman, G.; Arnaldsson, A.; Jónsson, H. A Fast and Robust Algorithm for Bader Decomposition of Charge Density. *Comput. Mater. Sci.* **2006**, *36* (3), 354–360. https://doi.org/10.1016/j.commatsci.2005.04.010.

34. Yu, M.; Trinkle, D. R. Accurate and Efficient Algorithm for Bader Charge Integration. *J. Chem. Phys.* **2011**, *134* (6), 064111. https://doi.org/10.1063/1.3553716.

35. Zhang, M.; Shao, X.; Liu, L.; Xu, X.; Pan, J.; Hu, J. 3d Transition Metal Doping Induced Charge Rearrangement and Transfer to Enhance Overall Water-Splitting on $Ni_3S_2$ (101) Facet: A First-Principles Calculation Study. *RSC Adv.* **2022**, *12* (41), 26866–26874. https://doi.org/10.1039/D2RA04252E.

36. Yan, Q. Q.; Wu, D. X.; Chu, S. Q. et al. Reversing the Charge Transfer Between Platinum and Sulfur-Doped Carbon Support for Electrocatalytic Hydrogen Evolution. *Nat. Commun.* **2019**, *10*, 4977. https://doi.org/10.1038/s41467-019-12851-w.

37. Hammer, B.; Nørskov, J. K. Electronic Factors Determining the Reactivity of Metal Surfaces. *Surf. Sci.* **1995**, *343* (3), 211–220. https://doi.org/10.1016/0039-6028(96)80007-0.

38. Di Liberto, G.; Cipriano, L. A.; Pacchioni, G. Universal Principles for the Rational Design of Single Atom Electrocatalysts? Handle with Care. ACS Catal. 2022, 12 (11), 6636–6645. https://doi.org/10.1021/acscatal.2c01011